\documentclass{aastex}
\newcommand{\be}{\begin{equation}}
\newcommand{\ee}{\end{equation}}
\begin{document}
\title{Magnetic Helicity Conservation and Astrophysical Dynamos}
\author{Ethan T. Vishniac}
\affil{Dept. of Physics and Astronomy, Johns Hopkins University, 
3400 N. Charles St., Baltimore MD 21210}
\email{ethan@pha.jhu.edu}

\and

\author{Jungyeon Cho}
\affil{Dept. of Astronomy, University of Wisconsin, 475 N Charter Street,  
Madison WI 53706}
\email{cho@astro.wisc.edu}

\begin{abstract}
We construct a magnetic helicity conserving dynamo theory 
which incorporates a calculated magnetic
helicity current.  In this model the fluid helicity plays a small
role in large scale magnetic field generation.
Instead, the dynamo process is dominated by a new quantity, 
derived from asymmetries in the 
second derivative of the velocity correlation function, closely
related to the `twist and fold' dynamo model. 
The turbulent damping term is, as expected, almost
unchanged. Numerical simulations with a spatially constant 
fluid helicity and vanishing resistivity are not
expected to generate large scale fields in equipartition
with the turbulent energy density.  The prospects
for driving a fast dynamo under these circumstances
are uncertain, but if it is possible, then the field
must be largely force-free. 
On the other hand, there is an efficient analog to
the $\alpha-\Omega$ dynamo.  Systems whose turbulence 
is driven by some anisotropic local instability in a shearing 
flow, like real stars and accretion disks, and some
computer simulations, may successfully drive
the generation of strong large scale magnetic fields, provided
that $\partial_r\Omega\langle \partial_\theta v_z\omega_\theta\rangle>0$.
We show that this criterion is usually satisfied.
Such dynamos will include a persistent, spatially coherent
vertical magnetic helicity current with the same sign as
$-\partial_r\Omega$, that is, positive for an accretion disk
and negative for the Sun.
We comment on the role of random magnetic helicity currents
in storing turbulent energy in a disordered magnetic field,
which will generate an equipartition, disordered field in
a turbulent medium, and also a declining long wavelength tail to
the power spectrum.  As a result, calculations of the 
galactic `seed' field are largely irrelevant.
\end{abstract}
\section{Introduction}

Astrophysical dynamos are usually
discussed in terms of mean-field dynamo theory
\citep[see][]{M78,P79,KR80}.
This typically involves several different assumptions.  First,
one assumes a dynamical separation between large scale and
small scale (i.e. turbulent eddy scale) fields.  Second, the
latter are assumed to be characterized by the turbulent
velocity field acting on the large scale field without
systematic velocity field effects due to the large scale field
(no `back-reaction').  Third, the large scale field is assumed to be smoothed by
turbulent diffusion.  (This is follows from the preceding
point and the assumption that reconnection is fast and efficient.)
Finally, the velocity field properties are typically prescribed
a priori, even when they are driven by magnetic field instabilities.
In this theory, the growth rate of the large scale magnetic field 
is driven by the fluid helicity.
Only the first point is a necessary part of dynamo theory.  In
fact, all other parts of mean-field dynamo theory have serious
problems 
\citep[see, for example,][]{CV91, P92, KA92, GD94, GD96, CH96, B00}.  
Obviously, the evidence for widespread dynamo
activity in stars suggests that the problems here lie in our
understanding of physics rather than suggesting that fast
dynamos are impossible. 

We will not address the issue of fast reconnection here.
There is a large body of evidence indicating that magnetic fields
in astrophysical plasmas can reconnect at speeds approaching the
Alfven speed 
\citep[see][ and references contained therein]{D96,IIAW97}.
This may be understood in terms of field line
stochasticity  \citep{LV99}
although there are
competing explanations (Petschek 1957; Shay et al. 1998; see also
Kulsrud 2000 for a modified version of Petschek's theory).  In addition,
the tendency to prescribe the small scale velocity field can be
seen as a largely formal problem.  Even when the turbulence is
driven by magnetic instabilities, e.g. the Balbus-Hawley
instability in accretion disks \citep{BH91,HB91}, 
as long as the eddy scale is much smaller than the
large scale magnetic field scale the problem can still be divided
into two conceptual steps: solving for the properties of the
small scale turbulence, and understanding the generation of the
large scale field.

Here we will focus on a particular aspect of the problem of
magnetic back-reaction.  We begin by noting that magnetic
helicity, ${\bf A}\cdot{\bf B}$, is strictly conserved for resistive MHD, as 
we take the resistivity to zero.  In contrast, while fluid helicity
${\bf v}\cdot{\bf\omega}$, is also conserved in the limit of zero viscosity, the
dissipation of fluid helicity at small scales does not vanish
as $\nu\rightarrow 0$, so that fluid helicity is not even
approximately conserved in any real turbulent system.  Unfortunately,
standard mean-field dynamo theory does not reflect this
conservation law. There must be a correlated backreaction that
enforces it, and this effect is left out of the standard theory. 

Numerical simulations seem to confirm the notion that there
is a serious problem with mean field dynamo theory.  
Computer simulations of dynamos can be divided into two classes.  There
are simulations in which some local instability (convection, the Balbus-Hawley
instability etc.) is allowed to operate, and there are simulations in
which the turbulence is driven externally, usually in such a way as to
guarantee the presence of a net fluid helicity.  The former simulations
are often successful at generating large scale magnetic fields whose
energy density is at least as great as the turbulent energy density
\citep[e.g.][]{HB92,GR95,BNST95}.
The latter are less successful, in the sense that the energy
density of the large scale magnetic field is often relatively modest 
\citep[e.g.][]{M81,Ba00}.
In recent years numerical simulations 
\citep{CH96,HCK96,B00}
have been performed to test the validity of mean field dynamo theory and
the role of magnetic helicity conservation in suppressing dynamo effects.
These calculations have used a closed box containing with some sort of
forced heliacal turbulence.  Cattaneo \& Hughes (1996) were able to
show a strong suppression of the turbulent dynamo in the limit of
small resistivity.  Hughes, Cattaneo and Kim (1996) did find
dynamo action, defined as the growth of the total magnetic energy
density, but it did not depend on the fluid helicity.  Their
eddy scale was very close to their box size, so a clean separation 
between eddy scale and large magnetic fields was not possible.
Finally, Brandenburg (2000) produced an example of a dynamo in
a computational box, with forced heliacal turbulence.  By varying the
resistivity he was able to demonstrate a steep inverse correlation
between the dynamo growth rate and the conductivity.  Naively
extrapolating to astrophysical regimes suggests that magnetic
dynamos driven by fluid helicity would take enormous amounts of
time to grow.

This result has been anticipated by a series of analytic
arguments and computational studies  (e.g. Vainshtein and Cattaneo 1992; Gruzinov
and Diamond 1994; 1996) which point to a suppression of
the electromotive force associated with the fluid helicity.
There have been attempts to calculate this back
reaction under various approximation schemes (e.g. Field, Blackman and Chou
1999), but they involve an expansion in parameters that are
typically of order unity\footnote{The expansion of Field et al. involves
a parameter which is almost exactly one in numerical simulations 
(see Cho and Vishniac 2000)}.  This is not
surprising, since the problem is similar to, but more complicated than,
attempts to derive the detailed properties of fluid turbulence
analytically.  Vainshtein and Cattaneo (1992) (see also Gruzinov and Diamond
1994) have suggested, from fairly
basic considerations, that this back reaction should be overwhelmingly
strong as soon as the magnetic field reaches equipartition with the
surrounding turbulence on the dissipation scale and that it should
suppress the critical components of the fluid helicity tensor
(i.e. those that contribute to an electromotive force parallel to
${\bf B}$) on large scales.  Since this criterion is satisfied when
the large scale field is negligibly small, this
looks like a fairly powerful anti-dynamo argument.  It is consistent
with the numerical work showing dynamo suppression.

Here we explore the possibility that a new version of mean
field dynamo theory, modified to explicitly incorporate
magnetic helicity conservation, can be used to predict the
evolution of large scale magnetic fields in highly conducting fluids. 
In section 2 of this paper we derive a new set of dynamo equations
and apply them to simple dynamo models.
We find that not all simple dynamos are eliminated, or even suffer
reduced growth rates.  
In section 3 we discuss the role of
random velocities in building a disordered magnetic field, and
apply our results to the early evolution of the galactic
magnetic field.  In the final section of this paper we summarize
our results and discuss some of the broader implications of this
work.

\section{Magnetic Helicity Conserving Magnetic Field Evolution}

We start with the usual expression for magnetic field evolution
in ideal MHD.
\be
\partial_t{\bf B_T}={\bf\nabla}\times({\bf v}\times{\bf B_T}),
\label{flux}
\ee
where the subscript `T' denotes the total field.  In what
follows, lower case letters will stand for fluctuation
quantities so that ${\bf B}_T\equiv {\bf B}+{\bf b}$.
Defining the vector potential, ${\bf A}$, in the usual way
we can show that
\be
\partial_t({\bf A}_T\cdot{\bf B}_T)
={\bf\nabla}\cdot\left(({\bf v}\times{\bf B}_T+\nabla\Phi_T)\times{\bf A}_T\right),
\label{helcon}
\ee
where $\Phi_T$ is defined through
\be
\nabla^2\Phi_T=\nabla\cdot({\bf v}\times{\bf B}_T),
\ee
and we have used the gauge condition $\nabla\cdot{\bf A}_T=0$.

Equation (\ref{helcon}) guarantees global conservation
of magnetic helicity.  It is equally obvious that the gauge
dependence inherent in the definition of magnetic helicity
means that this conservation law is gauge dependent.  We
have chosen $\nabla\cdot{\bf A}_T=0$, but any gauge condition
that does not include eddy scale terms will give equivalent
results.  It is tempting to use 
\be
\partial_t{\bf A}_T\equiv{\bf v}\times{\bf B}_T
\ee
as our gauge condition instead, so that we can drop $\nabla\Phi_T$
from equation (\ref{helcon}), but this introduces a steady growth
of magnetic helicity on eddy scales and we would no longer
expect large scale helicity to be separately conserved.

Mean field dynamo theory is an attempt to follow the magnetic
field averaged over many eddies, without a detailed calculation
of individual eddy scale features.  To this end we rewrite
equation (\ref{flux}) as
\be
\partial_t{\bf B}={\bf\nabla}\times{\cal E}_{mf},
\label{flux1}
\ee
where 
\be
{\cal E}_{mf}\equiv \langle {\bf v}\times{\bf b}\rangle.
\ee
The usual approach is to calculate ${\cal E}_{mf}$
using general assumptions about the structure of the underlying
turbulence.  Assuming that velocities are correlated only when
they are in the same direction, and that the energy is distributed
roughly isotropically among the available modes we find that
\be
{\cal E}_{mf,i}=\alpha_{ij}B_j-\epsilon_{ijk}D\partial_j B_k,
\ee
where
\be
D\equiv {1\over3}<v^2>\tau_c,
\ee
\be
\alpha_{ij}=\epsilon_{ilm}\langle v_l\partial_j v_m\rangle\tau_c,
\ee
and $\tau_c$ is a velocity correlation time.  The tensor
$\alpha_{ij}$ is the fluid helicity tensor. (Sometimes the
$\tau_c$ is omitted from the definition.)  Its trace, without
the factor of $\tau_c$, is the fluid helicity.

The difficulty with this approach is that, in general, it does
not conserve magnetic helicity.
If we rederive equation (\ref{helcon}) following only
the dynamics of the large scale fields, as defined in equation (\ref{flux1}),
we find that 
\begin{eqnarray}
\partial_t({\bf A}\cdot{\bf B})&=
{\bf A}\cdot{\bf\nabla}
\times{\cal E}_{mf}
+{\bf B}\cdot({\cal E}_{mf}-\nabla\Phi)
\label{helcon1}
\\
&=2{\bf B}\cdot{\cal E}_{mf}
+{\bf\nabla}\cdot\left(({\cal E}_{mf}+\nabla\Phi)
\times{\bf A}\right).
\nonumber
\end{eqnarray}
On the other hand, we can calculate the evolution of the average
magnetic helicity density by averaging equation (\ref{helcon}), 
over eddy scales.  We obtain
\be
\partial_t{\bf A}\cdot{\bf B}={\bf\nabla}\cdot
\left(({\cal E}_{mf}+\nabla\Phi)\times{\bf A}
+\langle({\bf e}_{mf}+\nabla\phi)\times{\bf a}\rangle\right).
\label{helcon2}
\ee
Here we have assumed that 
\be
\langle {\bf A}_T\cdot{\bf B}_T\rangle={\bf A}\cdot{\bf B},
\label{hels}
\ee
which will be true for an eddy scale smaller than the typical field scale
by a factor of at least ${\bf b}^2/{\bf B}^2$ and assuming efficient
transfer of magnetic helicity between scales.  The latter assumption
is examined at length in the appendix.  We will consider the
effects of a non-zero eddy scale in the third section of this paper.

The second contribution to the magnetic helicity current is produced
by correlations on eddy scales.  We will refer to it as the
anomalous magnetic helicity current, ${\bf J}_H$ (with an implied
minus sign, since it appears on the right hand side of equation (\ref{helcon2})).
Any self-consistent evolution equation for the large scale
magnetic field has to satisfy both of the preceding equations.
This does not uniquely specify the correct expression for ${\cal E}_{mf}$,
but if we further specify that this expression should depend only on the
local value of ${\bf B}$, the helicity current source
term in equation (\ref{helcon2}), and the conventional expression for 
${\cal E}_{mf}$ then we are forced to choose
\be
{\cal E}_{mf}={\cal E}_{mf,\perp}
-{{\bf B}\over 2 B^2}{\bf\nabla}\cdot {\bf J}_H,
\label{ans1}
\ee
where ${\cal E}_{mf,\perp}$ is the electromotive force
perpendicular to the direction of the large scale magnetic field.  
We will assume that this quantity can be calculated in the conventional
manner.  In any case, the critical element for a successful mean field
dynamo is the component of the electromotive force along the 
direction of the large scale magnetic field, and this is uniquely
specified.  The form of ${\cal E}_{mf}\cdot{\bf B}$ given in equation
(\ref{ans1}) was first proposed by Bhattacharjee and Hameiri
(1986), who further 
constrained the form of $J_H$ by requiring that the energy
dissipated by small scale currents be balanced locally by the
energy put into the large scale field.  We will not invoke
this constraint here, since in typical astrophysical systems
the energy flow through the large scale magnetic 
field is a small fraction of the energy flow through the
turbulent cascade, and does not need to be balanced locally
with ohmic dissipation to conserve energy.
While this work has since been extended to explore the
weak and strong magnetic field limits, and the role of hyperresistivity
in magnetic field evolution (Bhattacharjee and Yuan 1995), it has
not yet been fully applied to the generation of astrophysical
magnetic fields.  We note that 
equation (\ref{ans1}) also has the effect of eliminating the turbulent
dissipation of currents aligned with the large scale magnetic field.
The implication is that force-free magnetic fields are not subject to
turbulent dissipation, but all others are.  Gruzinov and Diamond (1994) 
concluded from this that the lifetimes of large scale astrophysical magnetic 
fields against dissipation are not qualitatively altered by magnetic
helicity conservation.  Here we note that this argument leaves open
the possibility that force-free magnetic fields might be generated
even by a very slow dynamo process.  Despite this loophole, we can 
conclude that the fast generation of large scale 
magnetic fields depends on the generation of large scale helicity currents from
eddy-scale processes.

In order to evaluate the helicity current term, we take
\be
{\bf a}=({\bf e}_{mf}-\nabla\phi)\tau_c,
\ee
where, as before, $\tau_c$ is the eddy correlation time.
Then equation (\ref{ans1}) becomes
\be
{\cal E}_{mf}={\cal E}_{mf,\perp}
+{{\bf B}\over B^2}{\bf\nabla}\cdot
\left(\langle\nabla\phi\times {\bf e}_{mf}\rangle\tau_c\right),
\label{ans2}
\ee
(This equation can also be recovered from equation (20)
in Bhattacharjee and Yuan (1995) in the limit where the resistivity
goes to zero and the eddy size is assumed to be much smaller
than the scale of the mean magnetic field.)
The source term in equation (\ref{ans2}) 
can be evaluated explicitly by writing $\phi$
in terms of the Fourier transform of ${\bf e}_{mf}$.
We find
\begin{eqnarray}
\langle\nabla\phi\times {\bf e}_{mf}\rangle
&={1\over(2\pi)^3}\langle\int\left(\int ({\bf I}-\hat k\hat k)
e^{i{\bf k}\cdot{\bf r}}d^3{\bf k}\cdot {\bf e}_{mf}({\bf x}+{\bf r})
\right)d^3{\bf r}\times {\bf e}_{mf}({\bf x})\rangle\\ 
&=\int {d^3{\bf r}\over 4\pi r} \epsilon_{ijk}
\partial_k\partial_l\langle {\bf e}_{mf,j}({\bf x})
 {\bf e}_{mf,l}({\bf x}+{\bf r})\rangle,
\nonumber
\end{eqnarray}
where the partial derivatives are with respect to the components
of ${\bf r}$.
This expression can be further simplified by using the first-order smoothing
approximation
\be
{\bf e}_{mf}={\bf v}\times({\bf B}+{\bf b})-\langle{\bf v}\times{\bf b}\rangle
\approx {\bf v}\times{\bf B}
\ee
and ignoring the gradient of the mean magnetic field over
eddy scales\footnote{Turbulent dissipation effects come from including
the field gradient terms when their coefficients are non-zero for
isotropic turbulence.  No such terms appear here.}.  
The use of the first-order smoothing approximation is
somewhat controversial, but in this context it does not violate 
any basic conservation laws.  This expression neglects the random
component of ${\bf e}_{mf}$.  Its effects are discussed in the
next section.  We find that 
${\bf J}_H$ is 
\be
{\bf J}_H=-\langle\nabla\phi\times {\bf e}_{mf}\rangle\tau_c
=-\int {d^3{\bf r}\over 4\pi r} \epsilon_{lnm} B_k
B_l\partial_k\partial_m\langle v_i({\bf x})v_n({\bf x}+{\bf r})\rangle\tau_c
\label{final}
\ee

We can rephrase this quantity in terms of correlations between
derivatives of the velocity field if we are willing to accept a 
certain loss of precision.  Expanding $v_n(({\bf x}+{\bf r})$ around
${\bf r}=0$ we find that equation (\ref{final}) becomes 
\be
{\bf J}_H
\sim -\lambda_c^2 \epsilon_{lnm}B_k
B_l\langle v_i({\bf x})\partial_m\partial_kv_n({\bf x})\rangle
\tau_c,
\label{diff}
\ee
where $\lambda_c$ is some sort of angle-averaged eddy correlation length,
roughly the geometric mean of the two largest perpendicular correlation
lengths.  If we integrate this by parts we obtain
\be
{\bf J}_H
\sim -\lambda_c^2
\langle({\bf B}\cdot{\bf\omega})
({\bf B}\cdot{\bf\nabla}){\bf v})\rangle
\tau_c.
\label{diff2}
\ee
Clearly there is
no simple relation between this and the fluid helicity.  In particular,
a computer simulation which is tailored to give a uniform non-zero
fluid helicity will not normally produce a non-zero magnetic
helicity current.  However, there is an attractive physical interpretation
to equation (\ref{diff2}).  The anomalous magnetic helicity current is
proportional to the correlated product
of the gradient of the velocity field along the magnetic field lines
(a `fold') and the vorticity along the magnetic field lines
(a rotation).  If one repeats this to form a complete dynamo cycle, 
it is obviously closely related to the `twist and fold' dynamo
model first proposed by Vainshtein and Zel'dovich (1972).

As a simple example we can consider the generation of a magnetic
field in a differentially rotating flow, $\Omega(r)\propto r^{-q}$.  
Galaxies, accretion disks, and stars are all instances of this case
(although the use of cylindrical geometry is a bit suspect for stars).
This is conventionally explained
as an example of an `${\alpha}-\Omega$' dynamo, where
\be
\partial_t B_r=-\partial_z{\cal E}_{mf,\theta},
\label{rev}
\ee
and
\be
\partial_t B_\theta=-q\Omega B_r+\partial_z\left(D\partial_z B_\theta\right).
\ee
For $B_r$ a small, fixed fraction of $B_\theta$  
we can write equation (\ref{rev}) as
\be
\partial_tB_r\sim -\partial_z B_\theta^{-1}
\partial_z\left(2\lambda_c^2B_\theta^2\langle\omega_\theta{1\over r}
\partial_\theta v_z\rangle\tau_c\right)+\partial_z\left({B_r\over B_\theta} D\partial_z
B_\theta\right).
\ee
For homogeneous anisotropic turbulence this becomes
\be
\left[\partial_t-D\partial_z^2\right]^2B_r=4q\Omega\lambda_c^2\tau_c\langle\omega_\theta{1\over r}
\partial_\theta v_z\rangle \partial_z^2B_r.
\label{ao}
\ee
In other words, a successful dynamo in this system 
requires that $q\partial_{\theta}v_z$
be negatively correlated with $\omega_\theta$ and that
the growth rate exceed the turbulent diffusion rate.

It is useful to compare this result with the result of a
conventional,  fluid helicity driven, `$\alpha-\Omega$'
dynamo.  In this case, ignoring dissipation, we get a growth rate 
\be
\Gamma\sim \left(\Omega{\alpha_{\theta\theta}\over L}\right)^{1/2},
\ee
where $L$ is a  large vertical scale associated with the structure of
the magnetic field and/or the structure of the disk or star.  Since
$\alpha_{\theta\theta}$ is odd for a reversal of any coordinate
direction, we need symmetry breaking in all three directions 
before we expect a non-zero value.  The differential
shear, and the resulting Coriolis forces, breaks symmetry
in the $\hat r$ and $\hat\theta$ directions, while ensuring
symmetry under the transformation 
$(\hat r,\hat\theta)\rightarrow(-\hat r,-\hat\theta)$
(which also leaves $\alpha_{\theta\theta}$ unchanged).  We
need to appeal to some kind of background vertical structure
to provide symmetry breaking in the $\hat z$ direction
before we can expect a nonzero $\alpha_{\theta\theta}$.
If we assume that the typical
eddy velocity is $\sim \lambda_c/\tau_c$, then 
$\alpha_{\theta\theta}\sim v\lambda_c(\Omega\tau_c)/L\sim \lambda_c^2\Omega/L$ and
\be
\Gamma\sim \Omega{\lambda_c\over L},
\ee
for the weak shearing limit, $\Omega\tau_c\le1$.
In contrast $\langle\omega_\theta{1\over r}\partial_\theta v_z\rangle$
is unchanged for $\hat z\rightarrow -\hat z$, and will generally
be non-zero for turbulence in a shearing flow.  No vertical
structure, aside from the vertical scale of the magnetic field,
is necessary.  Using equation (\ref{ao}) to estimate the
dynamo growth rate we find that
\be
\Gamma\sim \Omega{\lambda_c\over L_B}.
\label{aog}
\ee
Here $L_B$ is the vertical scale of the magnetic field, which
may be considerably less than the background structure scale, depending
on the efficiency of turbulent damping.  
The implication is that the growth rate of stellar and disk dynamos is
at least as high in this model as it is in models which
ignore magnetic helicity conservation, provided that
the local turbulence supplies the correct sign of 
$\langle\omega_\theta{1\over r}\partial_\theta v_z\rangle$.

How likely is this?  Invoking incompressibility and integrating
by parts, we can rewrite
this quantity as:
\begin{eqnarray}
\langle\omega_\theta{1\over r}\partial_\theta v_z\rangle
&=\langle\partial_z v_z{1\over r} \partial_\theta v_r\rangle
-\langle\partial_r v_z{1\over r}\partial_\theta v_z\rangle\\
&=-\langle\partial_r v_z{1\over r}\partial_\theta v_z\rangle
-\langle\partial_r v_r{1\over r}\partial_\theta v_r\rangle
-\langle{1\over r}\partial_\theta v_\theta{1\over r} 
\partial_\theta v_r\rangle.
\nonumber
\end{eqnarray}
If we replace these derivatives with wavenumbers, then we have
\be
\langle\omega_\theta{1\over r}\partial_\theta v_z\rangle
\approx-k_rk_\theta\langle v_r^2+v_z^2\rangle-
k_\theta^2\langle v_rv_\theta\rangle.
\label{estimate}
\ee
The effect of shear is to make $k_rk_\theta$ positive for
a positive $q$.  In other words, the first term always has
the correct sign for driving a dynamo.  The second term
may have either sign, but will only be significant in the
presence of a strong angular momentum flux.  For the Balbus-Hawley
instability in an accretion disks, $q=3/2$ and
$\langle v_rv_\theta\rangle>0$.  Consequently, both terms
in equation (\ref{estimate}) are negative and a dynamo is
guaranteed for sufficiently large vertical domain size.
In stars with active convection zones, like the Sun, the
result is less obvious, but will still favor dynamo activity
as long as the angular momentum distribution in the star is
stationary, or nearly so.  One striking counter-example is
provided by simulations of the magnetic Kelvin-Helmholtz instability
(Hawley, Gammie \& Balbus 1996), in which they repeated their
simulations of a zero-flux azimuthal magnetic field embedded  
in an accretion disk, but turned off the centrifugal force
term.  In this
case the absence of a centrifugal force term leads to a large
{\it negative} $\langle v_rv_\theta\rangle$, through turbulent
mixing, and no significant dynamo activity was observed.

We can understand the dynamics of this modified version of the
`$\alpha-\Omega$' dynamo by comparison with the standard model.
In the standard picture the fluid helicity induces a systematic 
spiral twisting of the azimuthal field lines.  Turbulent smoothing
gives a radial field provided that the twisting process gives a vertical
gradient in the product of the magnetic field strength and the
pitch angle of the spiral twisting.  By contrast, the process
considered here works through the action of the 
correlation $\langle v_z\partial_\theta\omega_\theta\rangle$ (which
is related to the left hand side of equation (\ref{estimate})
through an integration by parts).  This corresponds to twisting a 
bundle of field lines
in one direction when the bundle is displaced in the positive
$\hat z$ direction, and twisting it in the opposite sense when 
the $\hat z$ displacement is reversed.  Turbulent smoothing
gives us a net local excess of magnetic helicity when this
process is not uniform, that is, when $J_H$ has a non-zero
divergence.  Subsequently the turbulent smoothing of
bundles of spiralling magnetic field lines proceeds as in the
standard model.  Near a maximum in $|B_\theta|$
the effect of reconnection and smoothing is to produce
a coherent $B_r$ which, for the appropriate sign of the correlation,
will reinforce the azimuthal field component via shearing effects.
This may seem like a rather awkward substitute for the analogous
effect in the standard model, in which azimuthal field lines are
distorted into spirals with a coherent helicity.  However, this
new model involves only motions which are clearly realizable through
a continuous distortions of the fields lines, except for the final step
of reconnection, which involves reattachment of adjacent field lines
of differing orientation. The standard picture involves
a systematic rearrangement of the topology of the field lines
within the bundle, where the attachments of field lines are
not switched, but rather are allowed to slip over the surface
of a plane extending through the bundle.  The difference can be
readily appreciated by anyone who has ever attempted to twist
a rubber torus.   This is the reason why the standard model involves
a violation of magnetic helicity conservation, whereas the
model explored in this paper does not.

This particular example of a successful dynamo has a curious 
feature.  The total magnetic helicity current is
\be
{\bf J}_{tot}={\bf J}_H-({\cal E}_{mf}+\nabla\Phi)\times{\bf A}
={\bf J}_H+({{\bf B}\over2B^2}\nabla\cdot{\bf J}_H)\times{\bf A},
\label{helflo}
\ee
where we have dropped $\nabla\Phi=\partial_z\Phi$ and ${\cal E}_{mf,\perp}$ 
since they are both zero in this model.  The dominant component of the
${\bf A}$ is 
\be
A_r=-{\partial_zB_\theta\over k_z^2}, 
\ee
where $k_z$ is the vertical wavenumber of the magnetic field. 
Using  equation (\ref{diff2}) we can rewrite equation (\ref{helflo})
as
\be
{\bf J}_{tot}=(-\lambda_c^2
\langle\omega_\theta{1\over r}\partial_\theta v_z\rangle)
\left(B_\theta^2+{1\over k_z^2}(\partial_zB_\theta)^2\right).
\ee
This is a spatially constant, vertical magnetic helicity current.
Its divergence is zero, since the magnetic field has zero magnetic
helicity, but its presence is a necessary part of the dynamo
process.  We can add oscillating pieces by making different gauge
choices, but the large scale average is gauge invariant.  In
a real system this flux will appear at the vertical boundaries
of the system.  Its sign is a function of $q$.  For an accretion
disk ${\bf J}_{tot}\cdot\hat z>0$.  For a successful dynamo with
$q<0$ (like the Sun), we expect a persistent 
negative magnetic helicity current. 
Despite the fact that the solar dynamo is undoubtedly more
complicated than this simple example, this effect is observed
\citep{H27,R41,S90,W92,PCM95,RK96,CHM99}.

Finally, while this simple model has no net accumulation of magnetic
helicity, this is partly due to assuming perfect coupling between
the long wavelength magnetic helicity components due to small scale 
structures, which we refer to here, and in the appendix, as $h$,
and the long wavelength magnetic helicity components due to the
large scale field itself, denoted $H$.  We can estimate the
size of the fluctuations in $h$ from equation (\ref{A9}), which
implies that the fluctuations will be independent of $B$, and
reduced from the maximum magnetic helicity which can be stored
on the eddy scale by a factor $\sim (k_z\lambda)^{-1}$.

On the other hand, if we consider turbulence without a large scale
shear, then the role formerly played by the fluid helicity is
played by $-B^{-2}\nabla\cdot{\bf J}_H$.  While this will be non-zero,
even for homogeneous turbulence, it is hard to imagine a situation
where it can dominate over the turbulent dissipation term.  In
general, the two terms will be of the same order.  The former
can be made larger only at the expense of having a short wavelength
associated with the global properties of the turbulence, producing
a rapidly oscillating electromotive force.  Moreover, the dissipation
term depends on the diagonal terms in the Reynolds tensor, which
will be at least as large as the asymmetric terms which drive
${\bf J}_H$.  While this seems to be a strong argument against
the existence of a viable analog to the
`$\alpha^2$' dynamo of conventional mean-field dynamo theory, there
is a possible loophole.  Since the magnetic helicity conservation
constraint suppresses turbulent dissipation for force-free magnetic
fields, it follows that a dynamo process which preferentially
generates such a field may be possible.  We will defer any further 
discussion of this point to a subsequent paper.

\section{Random Effects in MHD Turbulence}

We have assumed in this discussion that the individual eddies
are negligibly small and that the only effects worth considering
are those that are coherent over many eddies.  
In typical astrophysical systems, this is an exaggeration, and
it is possible to generate some
large scale field even in the absence of an effective large scale
fast dynamo of the kind discussed above.  We note that
simulations of MHD turbulence usually show a magnetic field whose
energy density is 10\% or more of the turbulent energy density
with a typical scale which is large fraction of the typical eddy
scale (Meneguzzi et al. 1981).  Evidently, the random turbulent motions within
eddies are capable of producing large fluctuations of the
magnetic helicity on eddy scales.  

Assuming that this is the
case, we can estimate the long wavelength tail of the magnetic
field power spectrum by balancing the systematic dissipation
of magnetic helicity inhomogeneities with their random generation
via equation (\ref{helcon2}).  (The dissipative term in this equation
is not immediately apparent because it is divided between the
two magnetic helicity current terms on the right hand side.)
Given a typical eddy scale, $\lambda$, with a velocity, $v$, and fluctuating
magnetic field, $b$, we get a linear growth rate for the mean
square magnetic helicity on some larger scale $\Lambda$ as follows:
\be
\partial_t \langle ({\bf A}\cdot{\bf B})^2\rangle\sim
\Lambda^{-2} \langle v^2\rangle^2 \langle b^2\rangle^2\left({\lambda\over\Lambda}\right)^3
\tau_c^3,
\ee
where the factor of $(\lambda/\Lambda)^3$ comes from considering the fraction of
phase space containing the large scale modes.
Since the dissipation rate is just $\Lambda^{-2}\langle v^2\rangle\tau_c$,
we find that
\be
\langle ({\bf A}\cdot{\bf B})^2\rangle\sim 
\langle v^2\rangle\tau_c^2 \langle b^2\rangle^2\left({\lambda\over\Lambda}\right)^3.
\label{binput}
\ee
In other words, on large scales the magnetic helicity will settle into
a Poisson distribution after a dissipation time for the scale in question.
This implies that 
\be
B_\Lambda\sim b \left({\lambda\over \Lambda}\right)^{5/4},
\label{lfield}
\ee
which is small, but not necessarily negligible. 

Let's consider instead the situation where the magnetic field is in equipartition
with the turbulence on some very small scale, much smaller than the
typical energy-bearing eddy size.  That is, we consider $\lambda$ and
$\Lambda$ both small.  In this case we need to replace the turbulent
diffusion coefficient $\sim \langle v^2\rangle_\lambda^{1/2}\lambda$
with $\sim \langle v^2\rangle_\Lambda^{1/2}|\Lambda$.  
Assuming a Kolmogorov spectrum for $\langle v^2\rangle$,
this implies a turbulent diffusion coefficient which is
larger by a factor $(\Lambda/\lambda)^{4/3}$.  Nevertheless,
equation (\ref{lfield}) is still a fair estimate of the
strength of the magnetic field on the scale $\Lambda$ induced by random
anomalous magnetic helicity currents, since we should also
replace $\langle v^2\rangle_{\lambda}^2$ on the right hand
side of equation (\ref{binput}) with
$\langle v^2\rangle_{\Lambda}^2$, which is larger by the
same factor. In addition, we need to
consider the action of scales intermediate between $\lambda$
and $\Lambda$, where the magnetic field power spectrum is described by 
equation (\ref{lfield}).  As we consider larger values of $\lambda$
the term $\langle b^2\rangle^2$ in equation (\ref{binput}) declines
as $\lambda^{-5}$.  On the other hand, since these scales are part
of the inertial range of hydrodynamic turbulence, the factor 
$\tau_c^3$ scales as $\lambda^{2}$.  Including the factor 
of $(\lambda/\Lambda)^3$ as well we see that the magnetic helicity
currents driven on all intermediate scales, and the self-interaction
of fields on the scale $\Lambda$,  are as important
as the small scale driving force.  We conclude that the field
strength will increase exponentially at a rate which is comparable
to the eddy turn over rate on the scale of equipartition, until
the peak in the magnetic power spectrum approaches the
peak in the turbulent kinetic energy power spectrum.  This
is consistent with the numerical simulations cited above, and
lends support to the assumption of approximate equipartition 
between the magnetic field and the velocity field in a highly
conducting turbulent medium.

This has some interesting implications for the early evolution
of the galactic magnetic field.  If we imagine that the early
galactic disk was turbulent in roughly the same way the
it is today, that is with a typical fluid velocity of $10$ km/sec,
on scale $\sim 100$ pc, then we expect a disordered, equipartition
magnetic field to be present after a few times $10^7$ years.
The diffusion time for length scales of a few hundred parsecs
is an order of magnitude larger than the eddy turn over time
at $100$ pc, or about $3\times 10^8$ years, at which point we expect
the long wavelength tail to extend to magnetic domains with
sizes comparable to a disk thickness.  We are most concerned
with annular domains, which resist shearing and can serve
as the basis for generating a globally organized galactic
magnetic field.  These are oddly shaped, and we need to
rewrite the scaling law in equation (\ref{lfield})
as 
\be
B_{seed}\sim {b\over N_{eddy}^{1/4}} \left({\lambda\over \Lambda}\right)^{1/2},
\label{lfield2}
\ee
where $N_{eddy}$ is the number of independent turbulent eddies
in a magnetic domain.  This implies an annular magnetic field
of about $10^{-1} b$, or $\sim 3\times 10^{-7}$ gauss.  The
galactic dynamo operates at an efficiency which is some fraction
of the rotation speed divided by the galactic radius, or a
fraction of
$10^{-15}\hbox{sec}^{-1}$.  Even if the efficiency is as low
as $10$\% the large scale field will reach equipartition in
less than $10^9$ years.  The typical domain sizes should also continue
to increase, but a discussion of that process is outside the
scope of this paper.

We stress that while the physical processes invoked in this argument
are new, the basic idea of a self-generated turbulent field
as a critical step in the growth a galactic magnetic field
is not new.  The idea of generating a disordered field via
turbulent fluctuations can be found in Kulsrud, Cen, Ostriker
\& Ryu (1997), although
there the intent was to do without a large scale dynamo altogether.  
The physics of random {\it fluid} helicity fluctuations in a 
turbulent shearing medium has been previously discussed in Vishniac
\& Brandenburg (1997).  In addition, the early
growth of a galactic seed field, using the cross-helicity effect, 
has been discussed by Brandenburg \& Urpin (1998) and 
Blackman (2000).  This
model differs in producing exponential growth of an arbitrarily
weak small scale field using dynamo
processes which respect magnetic helicity conservation.
This growth proceeds through a kind of two-step inverse
cascade.  Large scale fluctuations in magnetic helicity
are generated by the spontaneous appearance of regions of 
positive and negative magnetic helicity.  These magnetic
helicity densities are then transfered to large scale
structures through dissipative coupling between the scales.

\section{Discussion and Conclusions}

We have examined the suppression of the dynamo effect due to 
the conservation of magnetic helicity.  This has a substantial
effect on the viability of dynamos operating in a closed box filled
with homogeneous turbulence, the `$\alpha^2$ dynamo'. 
Whether or not is completely suppresses them depends on whether or
not there is a version of our model which generates a largely
force-free dynamo field.  On the other hand, it has
remarkably little effect on the generation of large scale magnetic
fields in differentially rotating systems.  Dimensional
estimates of dynamo growth rates are the same; the need for
fully three dimensional turbulence to drive dynamo activity
is unchanged; and the important role of differential rotation
in producing large scale fields is unchanged.  Quantitatively,
we can see some differences.  The most obvious one is that
a previously ignored property of anisotropic turbulence 
plays a critical role in dynamo activity, and that there
is a strong tendency for this quantity to have the correct
sign for promoting dynamo activity.  A more subtle effect is
that the symmetry breaking requirements for a successful dynamo
are significantly reduced.  Finally, we note that a 
successful dynamo appears to rely on a large scale, spatially constant,
magnetic helicity current.  This current is not tied to
the magnetic domain size, and its sign depends only on the
sense of the differential rotation.

We have also examined
the role of random helicity currents in generating small
scale magnetic fields.  We find that this process plays a key
role in allowing magnetic fields to reach equipartition levels
in a turbulent medium.  Using plausible numbers for turbulence
in the galactic disk, we find that large scale magnetic fields
can be generated at levels of $\sim 10^{-1}$ of current
values within the first few $10^8$ years of the disk's existence,
starting from very small scale, highly disordered and
weak initial fields.  The strength of the galactic scale seed field
is irrelevant.

It seems odd that eddy scale motions can generate
significant large scale magnetic helicity currents when only 
a small amount of the large scale
helicity is contained in the eddies.  However,
while the eddies cannot store any significant amount of magnetic
helicity of one sign, they will create local fluctuations in the
magnetic helicity in the presence of a large scale field, 
even when there is no large scale magnetic helicity.
The typical size of these fluctuations will be 
$\sim Ba\sim B^2 v\tau_c$.  As long as these fluctuations can 
be moved in different directions, depending on their sign,
there is no paradox inherent in our results.  In particular
we note that the anomalous magnetic helicity is of the
same order as $v$ times the typical amplitude of these fluctuations.

We also note that while the usual mean-field dynamo violates magnetic helicity
conservation for the large scale field, this is not equivalent
to violating magnetic helicity conservation altogether.  The
total magnetic helicity that can be stored in individual eddies
is down from the magnetic helicity contained in a large scale
field with the same amplitude by a factor of $\lambda_c/L$.  The
implication is that the back-reaction which suppresses 
${\cal E}_{mf}\cdot{\bf B}$ won't set in until 
\be
B_L\sim v \left({\lambda_c\over L}\right)^{1/2}.
\ee
This should be unimportant when a fast dynamo of the
kind discussed in the second section of this paper is operating.  However,
it may be important otherwise, both in astrophysical objects
and in computer simulations.  Of course, if a successful dynamo
occurs, and creates a significant magnetic helicity density, then
this may generate secondarily a kinetic helicity.  We do not
expect this effect to play an important role in the dynamo process.

Finally, we have ignored the possibility that boundary conditions
might play a role in the evolution of the magnetic field.  
Blackman and Field (2000a,b,c) have suggested that the ejection of
magnetic flux across system boundaries could be a necessary
part of the dynamo process (see also Kleeorin et al. 2000).  
This would appear to conflict with the results of `shearing-box' simulations
of accretion disks (e.g. Hawley, Gammie \& Balbus\ 1996, and Hawley\ 2000) which
show the generation of a large scale field from initial conditions
with no net magnetic flux and little large scale structure, even with
periodic vertical boundary conditions.  Our results confirm that the
ejection of magnetic helicity is
not a requirement for a successful dynamo.
However, our results {\it do} imply that a large scale magnetic
helicity current is a necessary part of the dynamo process, and
in real systems (which typically lack periodic boundary conditions)
this will usually lead to a magnetic helicity current across the
system boundaries.  A non-periodic computer simulation with closed
boundaries will start to quench dynamo growth in a box of size $L$ when
the anomalous magnetic helicity current $\sim B^2 \langle v^2\rangle\tau_c$,
or equivalently $\sim B^2 D_T$, that accumulates in a magnetic domain
at the boundary becomes comparable to the magnetic helicity such a
domain can contain, or $\sim B^2 L_B$.  This is a quenching rate of
$D_T/L_B^2$, that is, the turbulent diffusion rate for a magnetic 
domain.  In a real system the magnetic helicity will be ejected
in a wind.  If the wind dimension is $L_W$ (presumably roughly the size
of the system) then since the ejected magnetic helicity flux must
be similar to the magnetic helicity flux required to run the dynamo
we expect an energy flux in the wind from a slow rotator of at least
\be
B_W^2 V_w\sim {\langle v^2\rangle \tau_c\over L_W} B^2\sim 
{\Gamma^2 B^2 L_B^2\over\Omega L_W}\sim B^2 L_B\Gamma\left({\lambda\over L_W}\right),
\ee
where we have used equation (\ref{aog}).  This energy flux is less
than the energy available from a single magnetic domain in the dynamo
system by the ratio of the eddy scale to the wind scale.  The magnetic
flux loss is bounded by $B_WV_W$, or
\be
B_WV_W\sim BL_B\Gamma \left({V_W\over\Omega L_W}\right).
\ee
For a slow rotator this limit can be more than the magnetic flux generation
rate within a single magnetic domain.  The implication is that either
the wind contains flux elements of mixed sign, or the wind is slow compared
to the escape velocity, or the wind becomes fast, but only at a distance
comparable to $V_{esc}/\Omega$.  

\acknowledgements We are happy to acknowledge helpful discussions
with Amitava Bhattacharjee, Eric Blackman, Patrick Diamond, Eun-Jin Kim, 
and Alex Lazarian.

\appendix

\section{The Transfer of Magnetic Helicity Between Scales}

Rather than assume that all long wavelength variations in
the magnetic helicity are equivalent, we could keep track of
the magnetic helicity due to large scale field components,
$H\equiv{\bf A}\cdot{\bf B}$, and that due to small scale field
components, $h\equiv\langle{\bf a}\cdot{\bf b}\rangle$, separately.
In this case, $J_H$ is clearly associated with small
scale structures, since it arises from eddy scale fluctuations
in the electromotive force. We can write the equations for $h$ 
and $H$ as
\be
\partial_t h=-\nabla\cdot{\bf J}_H-2{\bf B}\cdot\langle{\bf v\times b}\rangle_t,
\label{eq1}
\ee
and
\be
\partial_t H=\nabla\cdot(\langle {\bf v\times B}\rangle{\bf \times A}\rangle)
+2{\bf B}\cdot\langle{\bf v\times b}\rangle_t,
\label{eq2}
\ee
where the subscript `t' denotes effects which transfer magnetic helicity
between scales rather than between locations, and the sum of equations
(\ref{eq1}) and (\ref{eq2}) is equation (\ref{helcon2}).  
The standard mean field electromotive force parallel to ${\bf B}$
falls into the category of scale transfer effects.  However, it is
calculated by neglecting any information flow from small scales to
large, and consequently is not the entire term.  
Here we will show that the presence of a non-zero $h$
leads to the efficient transfer of magnetic helicity between
scales, which justifies lumping $H$ and $h$ together.

If $h$ is non-zero, then there is a piece of ${\bf a}$ which is
correlated with ${\bf b}$.  We can write this as
\be
{\bf a}_c=h{{\bf b}\over\langle b^2\rangle},
\ee
where we adopt the same gauge choice ($\nabla\cdot {\bf a}=0$) used
elsewhere in this paper.
This in turn implies a correlation between the small scale magnetic
field and the current,
\be
{\bf b}_c=h{{\bf \nabla\times b}\over\langle b^2\rangle},
\ee
where we have used the fact that the length scale for $h$ is much 
larger than
an eddy scale.  This extra piece of the magnetic field produces a 
first order change in the turbulent velocity field
\be
{\bf v}_c\approx {({\bf B\cdot\nabla}){\bf \nabla\times b}\over 4\pi\rho\langle b^2\rangle}h\tau_c,
\ee
which in turn produces a magnetic helicity transfer term of
\be
-2{\bf B\cdot}\langle{\bf v\times b}\rangle_t\approx
-2{h\tau_c\over4\pi\rho\langle b^2\rangle}
{\bf B\cdot}\langle ({\bf B\cdot\nabla}){\bf b\times}({\bf\nabla\times b})\rangle.
\ee
If the turbulence is approximately random and isotropic, then this reduces to
\be
-2{\bf B\cdot}\langle{\bf v\times b}\rangle_t\approx-2h\tau_cV_A^2k_{\|}^2
{b_{\perp}^2\over\langle b^2\rangle}\approx -{4\over9}h\tau_ck^2V_A^2,
\label{trate}
\ee
where the directions $\perp$ and $\|$ are defined relative to the direction
of the large scale field.

Equation (\ref{trate}) implies that small scale eddies dump their average 
magnetic helicity
into large scale structures at a rate which approaches the eddy turn over
rate as the large scale field approaches equipartition with the turbulence.
This expression does not include the back reaction of the large scale
field, which will be important if it already contains as much magnetic helicity as
possible on energetic grounds.  We will ignore this point here.  It may
have a quantitative effect on the saturation level of the magnetic field,
although it is unlikely to be more important than turbulent dissipation.
The transfer of magnetic helicity to larger scales can be treated as arbitrarily
fast provided that the dynamo growth rate, $\Gamma_D$, satisfies the
inequality
\be
\Gamma_D< \tau_c (kV_A)^2\sim {V_A^2\over\langle v^2\rangle}\tau_c^{-1}.
\ee
The implication is that for sufficiently weak large scale fields the
dynamo will start off by creating large scale gradients in the small
scale magnetic helicity, and subsequently evolve to the point where the
magnetic helicity is transferred to larger scales as fast as it accumulates.  This
will set in when the transfer term given in equation (\ref{trate}) is
roughly equal to $\nabla\cdot {\bf J}_H$ or
\be
h\sim {4\pi\rho\langle v^2\rangle\over k^2L_B},
\label{A9}
\ee
where $L_B$ is the length scale of the large scale magnetic field.  This
limit is the saturation value for the magnetic helicity on a scale $k^{-1}$
times $(kL_B)^{-1}$.  
In other words, the distinction between $h$ and $H$ 
will become dynamically unimportant long before $h$
approaches its saturation value.

In the presence of a non-zero fluid helicity we will also have 
the transfer of magnetic helicity between scales following the
standard mean field dynamo formalism.  This rate is comparable to
$\nabla\cdot J_H$ if we identify the symmetry breaking scale for
the fluid helicity with the length scale of the large scale magnetic
field. Consequently, when the large scale field is weak enough that
the transfer of small scale helicity to large scale helicities constitutes
an impediment for the dynamo process, the standard mean field dynamo
will be operating.  Conversely, once the large scale field is strong
enough to allow for efficient transfer of magnetic helicity between
scales, the fluid helicity is a subdominant effect.

Finally, in the absence of any large scale field at all, turbulent
dissipation alone will transfer magnetic helicity to large scales
by wiping out the smaller scale structures that generated the
large scale variations in magnetic helicity.  This process will
move helicity to larger scales, at the dissipation time scale
of the structures that contain the magnetic helicity fluctuations, until
the containing structures are as large as the length scale for the
magnetic helicity fluctuations.  Since this will go to completion on
the dissipation time scale for the largest scales involved, it is,
by definition, too slow to dominate in successful dynamos.  However,
it will be important in the kind of stochastic processes described
in section 3.  It dominates over the helicity transfer rate given
in equation (\ref{trate}) when 
\be
{\langle v^2\rangle \over L_B^2}v^2\tau_c> k^2V_A^2\tau_c
\ee
or
\be
V_A< \langle v^2\rangle^{1/2} {1\over kL_B}.
\ee


\begin{thebibliography}{}
\bibitem[Balbus \& Hawley\ 1991]{BH91}Balbus, S.A., \&
Hawley, J.F.\ 1991, \apj, 376, 214
\bibitem[Balsara\ 2000]{Ba00}Balsara, D.\ 2000, RevMexAA, 9, 92
\bibitem[Bhattacharjee \& Hameiri\ 1986]{BH86}Bhattacharjee, A. \&
Hameiri, E.\ 1986, \prl, 57, 206
\bibitem[Bhattacharjee \& Yuan\  1995]{BY95}Bhattacharjee, A. \&
Yuan, Y.\ 1995, \apj, 449, 739 
\bibitem[Blackman\ 2000]{Bl00}Blackman, E.G.\ 2000, \apj, 529, 138 
\bibitem[Blackman \& Field\ 2000a]{BF00a}Blackman, E.G., \& Field, G.B.\ 2000a, 
\apj, 534, 984
\bibitem[Blackman \& Field\ 2000b]{BF00b}Blackman, E.G., \& Field, G.B.\ 2000b, 
submitted to \mnras (astro-ph/9912459)
\bibitem[Blackman \& Field\ 2000c]{BF00c}Blackman, E.G., \& Field, G.B.\ 2000c, 
astro-ph/0009355
\bibitem[Brandenburg 2000]{B00}Brandenburg, A.\ 2000, astro-ph/0006186
\bibitem[Brandenburg et al.\ 1995]{BNST95}
Brandenburg, A., Nordlund, \AA, Stein, R.F., \& Torkelsson, U.\
1995, \apj, 446, 741
\bibitem[Brandenburg \& Urpin\ 1998]{BU98}Brandenburg, A. \& Urpin, V.
\ 1998, \aap, 332, L41
\bibitem[Canfield et al.\ 1999]{CHM99}Canfield, R. C., 
Hudson, H. S., \& McKenzie, D. E.\ 1999, Geophys. Res. Lett., 26, 627
\bibitem[Cattaneo \& Hughes\ 1996]{CH96}Cattaneo, F., \& Hughes, D.W.\ 1996,
\pre, 54, 4532
\bibitem[Cattaneo \& Vainshtein\ 1991]{CV91}Cattaneo, F., \& Vainshtein, S.I.\ 1991,
\apj, 376, L21
\bibitem[Cho \& Vishniac\ 2000]{CV00}Cho, J. \& Vishniac, E.T.\ 2000, submitted to \apj
\bibitem[Dere\ 1996]{D96} Dere, K.P. 1996, ApJ, 472, 864
\bibitem[Field et al.\ 1999]{FBC99}Field, G.B., Blackman, E.G., and
Chou, H.\ 1999, \apj, 513, 638
\bibitem[Glatzmaier \& Roberts\ 1995]{GR95}Glatzmaier, G.A., and Roberts, P.H.
\ 1995, Nature 377, 203
\bibitem[Gruzinov \& Diamond\ 1994]{GD94}Gruzinov, A.V. \& Diamond, P.H.\ 1994,
\prl, 72, 1651
\bibitem[Gruzinov \& Diamond\ 1996]{GD96}Gruzinov, A.V. \& Diamond, P.H.\ 1996,
Phys. Plasmas, 3, 1853
\bibitem[Hale\ 1927]{H27}Hale, G. E.\ 1927, Nature, 119, 708
\bibitem[Hawley\ 2000]{H00}Hawley, J.F.\ 2000, astro-ph/0011501
\bibitem[Hawley \& Balbus\ 1991]{HB91}Hawley, J.F., \& Balbus, S.A.\ 1991, 
\apj, 376, 223
\bibitem[Hawley \& Balbus\ 1992]{HB92} Hawley, J.F. \& Balbus, S.A.\ 1992, \apj, 400, 610
\bibitem[Hawley, Gammie \& Balbus\ 1996]{HGB96}Hawley, J.F., Gammie, C.F \& Balbus, 
S.A.\ 1996,\apj, 464, 690
\bibitem[Hughes et al.\ 1996]{HCK96}
Hughes, D.W., Cattaneo, F. \& Kim, E.J.\ 1996, Phys. Lett. A, 223, 167
\bibitem[Innes et al.\ 1997]{IIAW97}
Innes, D.E., Inhester, B., Axford, W.I., \& Wilhelm, K. 1997, Nature, 386, 811
\bibitem[Kleeorin, Moss, Rogachevskii, \& Sokoloff\ 2000]{KMRS00}
Kleeorin, N., Moss, D., Rogachevskii, \& Sokoloff, D.\ 2000, \aap, 361, L5
\bibitem[Krause \& Radler\ 1980]{KR80} Krause, F., \& Radler, K.H. 1980,
Mean-Field Magnetohydrodynamics and Dynamo Theory (Oxford: Pergamon Press)
\bibitem[Kulsrud\ 2000]{K00}Kulsrud, R.M.\ 2000, astro-ph/0007075
\bibitem[Kulsrud \& Anderson\ 1992]{KA92}
Kulsrud, R.M., \& Anderson, S.W. \ 1992, \apj, 396, 606
\bibitem[Kulsrud, Cen, Ostriker \& Ryu\ 1997]{KCOR97}Kulsrud, R.M.,
Cen, R., Ostriker, J.P. \& Ryu, D.\ 1997, \apj, 480, 481
\bibitem[Lazarian \& Vishniac\ 1999]{LV99}Lazarian, A., \& Vishniac, E.T.
\ 1999, \apj, 517, 700
\bibitem[Meneguzzi et al.\ 1981]{M81}
Meneguzzi, M., Frisch, U. and Pouquet, A., 1981, Phys. Rev. Lett., 47, 1060
\bibitem[Moffatt\ 1978]{M78} Moffatt, H.K. 1978, Magnetic Field Generation in E
lectrically
Conducting Fluids (Cambridge: Cambridge University Press)
\bibitem[Parker\ 1979]{P79}\rule{1.2cm}{0.2mm}\ 1979, Cosmical
Magnetic Fields (Oxford: Clarendon Press)
\bibitem[Parker\ 1992]{P92}Parker, E.N.\ 1992, \apj, 401, 137
\bibitem[Petschek\ 1964]{P64}Petschek, H.E.\ 1964,
{\it The Physics of Solar Flares}, AAS-NASA
Symposium, NASA SP-50 (ed. W.H. Hess), Greenbelt, Maryland, p.~425
\bibitem[Pevtsov et al.\ 1995]{PCM95}Pevtsov, A. A., 
Canfield, R. C., \& Metcalf, T. R.\ 1995, \apj, 440, L109 
\bibitem[Richardson\ 1941]{R41}Richardson, R. S.\ 1941, \apj, 41, 24
\bibitem[Rust \& Kumar\ 1996]{RK96}Rust, D. M., \& Kumar, A.\ 1996, 
\apj, 464, L199 
\bibitem[Seehafer\ 1990]{S90}Seehafer, N.\ 1990, Sol. Phys., 125, 219
\bibitem[Shay, Drake, Denton, \& Biskamp\ 1998]{SDDB98}Shay, M.A., Drake, J.F.,
Denton, R.E., \& Biskamp, D.\ 1998, J. Geophys. Res., 103, 9165
\bibitem[Vainshtein \& Cattaneo\ 1992]{VC92}Vainshtein, S.I. and Cattaneo, F.
\ 1992, \apj, 393, 165
\bibitem[Vainshtein \& Zel'dovich]{VZ72} Vainshtein, S.I. \& Zel'dovich, Ya. B.
\ 1972, Usp. Fiz. (SSSR) 106, 431 (Sov. Phys. Usp., 15, 159)
\bibitem[Vishniac \& Brandenburg\ 1997]{VB97}Vishniac, E.T. \&
Brandenburg, A.\ 1997, \apj, 475, 263
\bibitem[Webb\ 1992]{W92}Webb, D. F. 1992, 
in Eruptive Solar Flares, ed. Z. Svestka, 
B. V. Jackson, \& M. E. Machado (New York: Springer), 234
\end{thebibliography}
\end{document}